\documentclass[12pt]{article}
\usepackage[a4paper, total={7in, 10in}]{geometry}
\usepackage[parfill]{parskip}
\usepackage{physics, tensor, float, subcaption}
\usepackage{graphicx}
\graphicspath{ {Plots/} }
\usepackage{jhep-mod}
\usepackage{bm}
\usepackage{soul}
\usepackage{amssymb,amsmath,amsthm}
\usepackage{mathrsfs}
\usepackage[utf8]{inputenc}
\usepackage{enumerate}
\usepackage{bigints}
\usepackage{xcolor}
\usepackage{appendix}
\usepackage{graphicx}
\usepackage{float}
\usepackage{tikz}
\usepackage{setspace}
\usepackage{cancel}
\definecolor{purple}{rgb}{1,0,1}
\definecolor{lime}{HTML}{A6CE39} % needs xcolor
\definecolor{lime}{HTML}{A6CE39}
\newcommand{\orcidicon}{%
	\begin{tikzpicture}
	\draw[lime, fill=lime] (0,0) 
		circle [radius=0.16] 
		node[white] {{\fontfamily{qag}\selectfont \tiny ID}};
	\draw[white, fill=white] (-0.0625,0.095) 
		circle [radius=0.007];
	\end{tikzpicture}
	\hspace{-5mm}
}
\newcommand\orcidAlex{{\href{https://orcid.org/0000-0002-1763-3563}{\orcidicon}}}

\newcommand{\e}{\mathrm{e}}

\newcommand{\V}{\mathcal{V}}
\begin{document}
%========================================================

%========================================================
%========================================================

\title{\vspace{-25pt}\huge{
Ringing of the regular black hole with asymptotically Minkowski core
}}
%Old title: 
% Dynamic thin--shell wormholes constructed from surgery 
% of regular black holes with asymptotically Minkowski cores

%========================================================
%========================================================
%========================================================
\author{
\Large
Alex Simpson\!\orcidAlex\!
}
%========================================================
%========================================================
%========================================================
%========================================================
\affiliation{School of Mathematics and Statistics, Victoria University of Wellington, \\
\null\qquad PO Box 600, Wellington 6140, New Zealand.}
%========================================================
%========================================================
%\emailAdd{thomas.berry@sms.vuw.ac.nz}
\emailAdd{alex.simpson@sms.vuw.ac.nz}
%\emailAdd{matt.visser@sms.vuw.ac.nz}
%========================================================
%========================================================

\abstract{
\vspace{1em}

A Regge--Wheeler analysis is performed for a novel black hole mimicker `the regular black hole with asymptotically Minkowski core', followed by an approximation of the permitted quasi-normal modes for propagating waveforms. A first-order WKB approximation is computed for spin zero and spin one perturbations of the candidate spacetime. Subsequently, numerical results analysing the respective fundamental modes are compiled for various values of the $a$ parameter (which quantifies the distortion from Schwarzschild spacetime), and for various multipole numbers $\ell$. Both electromagnetic spin one fluctuations and scalar spin zero fluctuations on the background spacetime are found to possess shorter-lived, higher-energy signals than their Schwarzschild counterparts for a specific range of interesting values of the $a$ parameter. Analysis as to what happens when one permits perturbations of the Regge--Wheeler potential itself is then conducted, first in full generality, before specialising to Schwarzschild spacetime. A general result is presented explicating the shift in quasi-normal modes under perturbation of the Regge--Wheeler potential.

\bigskip

\bigskip
\noindent
{\sc Date:} 12 July 2021; \LaTeX-ed \today


\bigskip
\noindent{\sc PhySH:} 
Gravitation
}

%========================================================
\maketitle
%========================================================
\def\tr{{\mathrm{tr}}}
\def\diag{{\mathrm{diag}}}
\def\cof{{\mathrm{cof}}}
\def\pdet{{\mathrm{pdet}}}
\parindent0pt
\parskip7pt

%=====================================================
\section{Introduction}
%=====================================================

Given the conditions that a propagating waveform is purely ingoing at the horizon and purely outgoing at spatial infinity, the proper oscillation frequencies of a candidate black hole spacetime are determined \emph{via} analysis of the permitted quasi-normal modes (QNMs). QNM analysis is by now utterly standard, with a wealth of literature containing QNM analyses in many varied contexts~\cite{Vishveshwara:1970, Zerilli:1970, Press:1971, Davis:1971, Price:1971, Price:1972, Bardeen:1973, Zerilli:1974, Chandrasekhar:1975, Detweiler:1977, Detweiler:1979, Blome:1984, Ferrari:1984, Bachelot:1993, Fiziev:2006, Konoplya:2011, Bronnikov:2012, Aneesh:2018, Santos:2019, Aragon:2020, Cuadros:2020, Churilova:2019}, as well as the QNMs of propagating waveforms emanating from an astrophysical source being directly observed \emph{via} experiment in the LIGO/VIRGO merger events~\cite{ligo-detection-papers, grav-wave-observations-wiki}. Given the hope that LIGO/VIRGO (or more likely LISA~\cite{LISA}) will eventually be able to delineate the fingerprint of classical black holes from possible black hole mimickers, it is increasingly relevant to analyse well-motivated candidate spacetimes that model black hole mimickers and to compile results that speak to the advances made in observational and gravitational wave astronomy.

One such candidate spacetime is the regular black hole with an asymptotically Minkowski core. By `regular black hole', one means in the sense of Bardeen~\cite{Bardeen:1968}; a black hole with a well-defined horizon structure and everywhere-finite curvature tensors and curvature invariants. Regular black holes as a subject matter possess a rich genealogy; see for instance references~\cite{Bardeen:1968, Hayward:2005, Bronnikov:2005, Bambi:2013, Frolov:2014, Simpson:2019, Mazza:2021, Franzin:2021}. For current purposes, the candidate spacetime in question is given by the line element
\begin{equation}
\dd s^2 = -\left(1-\frac{2m\,\e^{-a/r}}{r}\right)\dd{t}^2 + \frac{\dd{r}^2}{1-\frac{2m\,\e^{-a/r}}{r}} + r^2\left(\dd{\theta}^2 + \sin^2\theta \dd{\phi}^2\right) \ .
\label{metric}
\end{equation}
One can find thorough discussions of aspects of this specific metric in both~\cite{Simpson:2020, Berry:2020}, where causal structure, surface gravity, satisfaction/violation of the standard energy conditions, and locations of both photon spheres and timelike circular orbits are analysed through the lens of standard general relativity. An extremal version of this metric, and various other metrics with mathematical similarities, have also been discussed in rather different contexts in~\cite{Culetu:2013, Xiang:2013, Culetu:2014, Culetu:2015a, Culetu:2015b, Junior:2015, Rodrigues:2015, Takeuchi:2016}.

This paper seeks to compute some of the relevant QNM profiles for this candidate spacetime. Consequently, the author first performs the necessary extraction of the specific spin-dependent Regge--Wheeler potentials in \S~\ref{S:RW_potential}, before analysing the spin one and spin zero QNMs \emph{via} the numerical technique of a first-order WKB approximation in \S~\ref{S:QNMs}. For specified multipole numbers $\ell$, and various values of $a$, numerical results are then compiled in \S~\ref{S:Results}. These analyse the respective fundamental modes for spin one and spin zero perturbations of a background spacetime possessing some trial astrophysical source. General perturbations of the Regge--Wheeler potential itself are then analysed in \S~\ref{S:RWperturb}, with some quite general results being presented, before concluding the discussion in \S~\ref{S:discussion}.

%=====================================================
\section{Regge--Wheeler potential}\label{S:RW_potential}
%=====================================================

In this section the spin-dependent Regge--Wheeler potentials are explored. Ultimately, the spin two axial mode involves perturbations which are somewhat messier, and hence do not lend themselves nicely to the WKB approximation and subsequent computation of quasi-normal modes without the assistance of numerical code. Due to this ensuing intractability, the relevant Regge--Wheeler potential for the spin two axial mode is explored for completeness, before specialising the QNM discourse to spin zero (scalar) and spin one (electromagnetic, \emph{e.g.}) perturbations only. The QNMs of spin two axial perturbations are relegated to the domain of future research. Given one does not know the spacetime dynamics \emph{a priori}, the inverse Cowling approximation is invoked, where one allows the scalar/vector field of interest to oscillate whilst keeping the candidate geometry fixed. This formalism closely follows that of~\cite{Boonserm:2013}.

To proceed, one implicitly defines the tortoise coordinate \emph{via}
\begin{equation}
\dd{r^*} = \frac{\dd{r}}{1-\frac{2m\,\e^{-a/r}}{r}} \ .
\label{tortoise}
\end{equation}
Although this equation is not analytically integrable, one can still conduct analysis of the Regge--Wheeler potential through this implicit definition of the tortoise coordinate. The coordinate transformation Eq.~\eqref{tortoise} allows one to write the spacetime metric Eq.~\eqref{metric} in the following form:
\begin{equation}
\dd{s}^2 = \left( 1- \frac{2m\,\e^{-a/r}}{r} \right) \bigg\{ -\dd{t}^2 + \dd{r_*}^2 \bigg\} + r^2 \left( \dd{\theta}^2 + \sin^2\theta \dd{\phi}^2 \right) \ ,
\end{equation}
which can then be rewritten as
\begin{equation}\label{metricstar}
\dd{s}^2 = A(r_*)^2 \big\{ -\dd{t}^2 + \dd{r_*}^2 \big\} + B(r_*)^2 \left( \dd{\theta}^2 + \sin^2\theta \dd{\phi}^2 \right) \ .
\end{equation}
In Regge and Wheeler's original work \cite{Regge:1957}, they show that for perturbations in a black hole spacetime, assuming a separable wave form of the type
\begin{equation}
\Psi(t,r_*,\theta,\phi) = \e^{i\omega t} \psi(r_*) Y(\theta,\phi)
\label{sepwaveform}
\end{equation}
results in the following differential equation (now called the Regge--Wheeler equation):
\begin{equation}
\pdv[2]{\psi(r_*)}{r_*} + \big\{ \omega^2 - \V_S \big\}\,\psi(r_*) = 0 \ .
\label{RWeq}
\end{equation}
Here \( Y(\theta,\phi) \) represents the spherical harmonic functions, \( \psi(r_*) \) is a propagating scalar, vector, or spin two axial bivector field in the candidate spacetime, \( \V_S \) is the spin-dependent Regge--Wheeler potential, and \( \omega \) is some (possibly complex) temporal frequency in the Fourier domain~\cite{Fiziev:2006, Aneesh:2018, Santos:2019, Regge:1957, Boonserm:2013, Simpson:2019, Boonserm:2018}. The method for solving Eq.~\eqref{RWeq} is dependent on the spin of the perturbations and on the background spacetime. For instance, for vector perturbations (\(S=1\)), specialising to electromagnetic fluctuations, one analyses the electromagnetic four-potential subject to Maxwell's equations:
\begin{equation}
    \frac{1}{\sqrt{-g}}\,\partial_{\mu}\left(F^{\mu\nu}\,\sqrt{-g}\right) = 0 \ ,
\end{equation}
whilst for scalar perturbations ($S=0$), one solves the minimally coupled massless Klein--Gordon equation
\begin{equation}
    \square\psi(r) = \frac{1}{\sqrt{-g}}\,\partial_{\mu}\left(\sqrt{-g}\,\partial^{\mu}\,\psi\right) = 0 \ .
\end{equation}
Further details can be found in references~\cite{Santos:2019, Aragon:2020, Regge:1957, Boonserm:2013}. For spins $S\in\lbrace0,1,2\rbrace$, this yields the general result in static spherical symmetry~\cite{Boonserm:2013, Boonserm:2018}:
\begin{equation}
\V_{0,1,2} = \left\{\frac{A^2}{B^2}\right\} \left[\ell(\ell+1)+S(S-1)(g^{rr}-1)\right] + (1-S) \frac{\partial^2_{r_*}B}{B} \ ,
\end{equation}
where $A$ and $B$ are the relevant functions as specified by Eq.~(\ref{metricstar}), $\ell$ is the multipole number (with $\ell\geq S$), and $g^{rr}$ is the relevant contrametric component with respect to standard curvature coordinates (for which the covariant components are presented in Eq.~(\ref{metric})).

For the spacetime under consideration, one has $A(r)=\sqrt{1-\frac{2m\,\e^{-a/r}}{r}}$, \( B(r) = r \), $g^{rr} = 1-\frac{2m\,\e^{-a/r}}{r}$, and \( \partial_{r_*} = \left(1-\frac{2m\,\e^{-a/r}}{r}\right) \partial_r \). Hence,
\begin{equation}
\frac{\partial^2_{r_*}B}{B} = \frac{\left(1-\frac{2m\,\e^{-a/r}}{r}\right)\partial_r\left[1-\frac{2m\,\e^{-a/r}}{r}\right]}{r} 
= \left( \frac{r - 2m\,\e^{-a/r}}{r^3} \right) \left( \frac{2m\,\e^{-a/r}(r-a)}{r^2} \right) \ ,
\end{equation}
and so one has the exact result that
\begin{equation}
\V_{0,1,2} = \left( \frac{r - 2m\,\e^{-a/r}}{r^3} \right) \left\{ \ell(\ell+1) + \frac{2m\,\e^{-a/r}}{r}(1-S)\left[S + 1 - \frac{a}{r}\right] \right\} \ .
\end{equation}
That is,
\begin{equation}
\V_{0,1,2} = \left( 1- \frac{2m\,\e^{-a/r}}{r} \right) \left\{ \frac{\ell(\ell+1)}{r^2}  + \frac{2m\,\e^{-a/r}}{r^3}(1-S)\left[S+1-\frac{a}{r}\right] \right\} \ .
\end{equation}
Note that at the outer horizon, \( r_H = 2m \,\e^{W\left(-\frac{a}{2m}\right)} \), with $W$ being the special Lambert $W$ function~\cite{Boonserm:2013, Boonserm:2018, Valluri:2000, Valluri:2009, Boonserm:2008, Boonserm:2010, Sonoda:2013a, Sonoda:2013b, Corless:1996, Vial:2012, Stewart:2011, Stewart:2012, Visser:2018-LW}, the Regge--Wheeler potential vanishes. Taking the limit as \( a\rightarrow0 \) recovers the known Regge--Wheeler potentials for spin zero, spin one, and spin two axial perturbations in the Schwarzschild spacetime:
\begin{equation}
\V_{Sch.,0,1,2} = \lim_{a\rightarrow0} \V_{0,1,2} = \left(1-\frac{2m}{r}\right) \left\{ \frac{\ell(\ell+1)}{r^2} + \frac{2m}{r^3}(1-S^2) \right\} \ .
\end{equation}
Note that in Regge and Wheeler's original work~\cite{Regge:1957}, only the spin two axial mode was analysed. However, this result agrees both with the original work, as well as with later results extending to spin zero and spin one perturbations~\cite{Santos:2019}. It is informative to explicate the exact form for the RW-potential for each spin case, and to then plot the qualitative behaviour of the potential as a function of the dimensionless variables $r/m$ and $a/m$ for the respective dominant multipole numbers ($\ell=S$).

\begin{itemize}
    \item Spin one vector field: The conformal invariance of spin one massless particles in $(3+1)$ dimensions implies that the $\frac{\partial_{r_*}^2B}{B}$ term vanishes, and indeed mathematically the potential reduces to the highly tractable
    \begin{equation}
        \V_1 = \left(1-\frac{2m\,\e^{-a/r}}{r}\right)\frac{\ell(\ell+1)}{r^2} \ .
    \end{equation}
    Specialising to the dominant multipole number $\ell=1$ gives:
    \begin{equation}\label{ell1}
        \eval{\V_1}_{\ell=1} = \frac{2}{r^2}\left(1-\frac{2m\,\e^{-a/r}}{r}\right) \ .
    \end{equation}
    Now, in order to examine the qualitative features of the potential it is of mathematical convenience to define the new dimensionless variables $x=r/m$, and $y=a/m$. It is worth noting here that convention in the historical literature would be to set $a\sim m_{p}$, such that the newly introduced scalar parameter appeals to the quantum gravity regime. This would imply that $y=a/m\sim m_{p}/m_{sun}\ll 1$. In view of the redefinition of parameters, Eq.~(\ref{ell1}) may be re-expressed as follows:
    \begin{equation}
        \eval{m^2\V_1}_{\ell=1} = \frac{2}{x^2}\left(1-\frac{2\,\e^{-y/x}}{x}\right) \ .
    \end{equation}
    The qualitative features of $\V_1$ are then plotted in Fig.~\ref{F:1}, for the full range of $y$ such that the spacetime still possesses a nontrivial horizon structure, and the domain for $x$ such that one is strictly outside the horizon.
    %
%=====================================================
\begin{figure}[htb!]
%\vspace{-1cm}
\begin{subfigure}{.45\textwidth}
\includegraphics[scale=0.32]{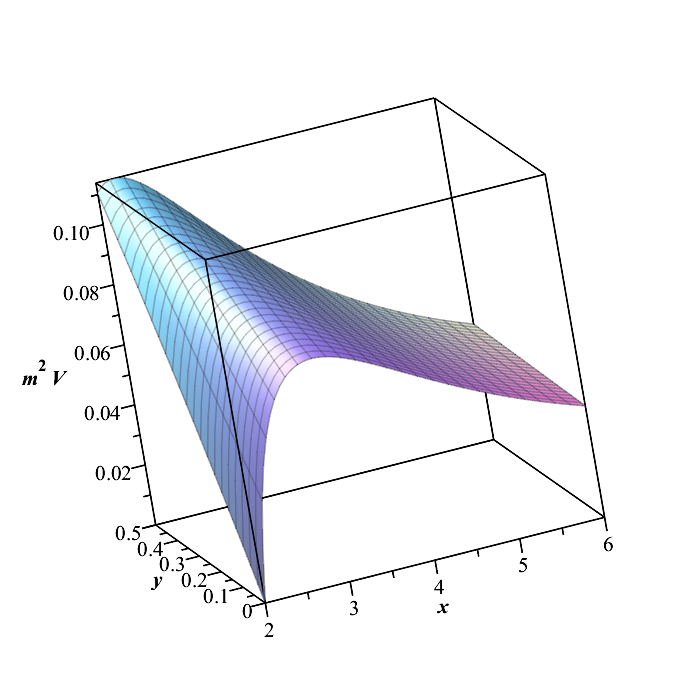}
\caption{Three-dimensional plot of $m^2\V_1$.}
\end{subfigure}
\begin{subfigure}{.45\textwidth}
\includegraphics[scale=0.36]{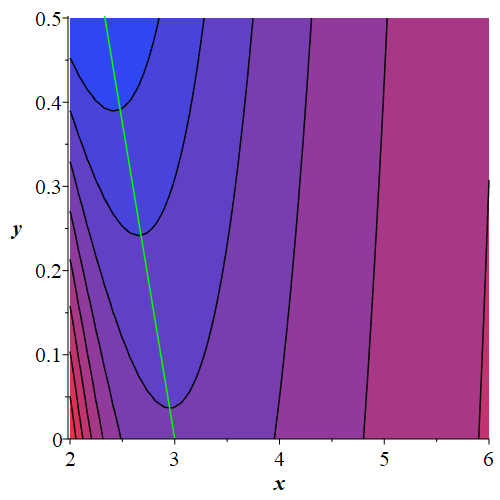}
\caption{Contour plot. `blue'$\rightarrow$`red' corresponds to `high'$\rightarrow$`low'.}\label{F:1b}
\end{subfigure}
\caption{The qualitative features of the spin one Regge--Wheeler potential for the dominant multipole number $\ell=1$ are depicted.}
\label{F:1}
\end{figure}
%=====================================================

    An immediate sanity check from Fig.~\ref{F:1} is that for $a=0$, where the candidate spacetime reduces to Schwarzschild, one observes a peak at $r=3m$. This is the expected location of the photon sphere for Schwarzschild, and is indeed the corresponding location of the peak of the relevant spin one RW-potential. As $a$ increases, the $r$-coordinate location of the peak decreases. For all values of $a$, there is falloff at spatial infinity, and once the peak is crested there is rapid falloff as one approaches the horizon location (where the RW-potential vanishes completely). The green line present in Fig.~\ref{F:1b} corresponds to the approximate location of the photon sphere calculated in reference~\cite{Berry:2020}; $r_\gamma\approx 3m-\frac{4}{3}a$. This approximation is used for the location of the peak of the spin one potential in order to extract the QNM profile approximations in \S~\ref{S:QNMs}. One can see that for the given domain and range this approximation has high accuracy, closely matching with the locations of the peaks.
    \item Spin zero scalar field: The potential now becomes
    \begin{equation}
        \V_0 = \left(1-\frac{2m\,\e^{-a/r}}{r}\right)\left\lbrace \frac{\ell(\ell+1)}{r^2}+\frac{2m\,\e^{-a/r}}{r^3}\left(1-\frac{a}{r}\right)\right\rbrace \ ,
    \end{equation}
    and, fixing the dominant multipole number $\ell=0$, one specialises to the scalar $s$-wave, which is of particular importance, yielding
    \begin{equation}
        \eval{\V_0}_{\ell=0} = \frac{2m\,\e^{-a/r}}{r^3}\left(1-\frac{2m\,\e^{-a/r}}{r}\right)\left(1-\frac{a}{r}\right) \ .
    \end{equation}
    Once again, to examine the qualitative features of the potential it is convenient to re-express this in terms of the dimensionless variables $x=r/m, \ y=a/m$:
    \begin{equation}
        \eval{m^2\V_0}_{\ell=0} = \frac{2\,\e^{-y/x}}{x^3}\left(1-\frac{2\,\e^{-y/x}}{x}\right)\left(1-\frac{y}{x}\right) \ .
    \end{equation}
    The qualitative features of $\V_0$ are then displayed in Fig.~\ref{F:2}.
    %
    %=====================================================
\begin{figure}[htb!]
%\vspace{-1cm}
\begin{subfigure}{.45\textwidth}
\includegraphics[scale=0.32]{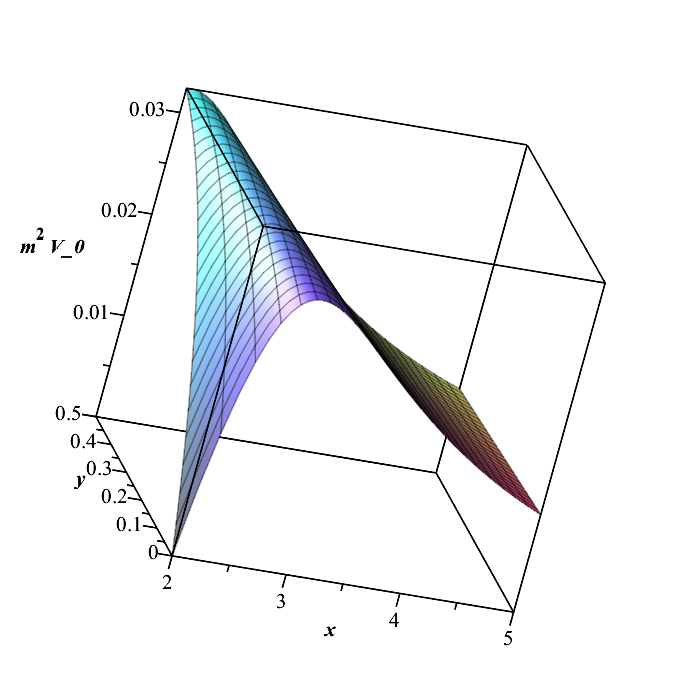}
\caption{Three-dimensional plot of $m^2\V_0$.}
\end{subfigure}
\begin{subfigure}{.45\textwidth}
\includegraphics[scale=0.36]{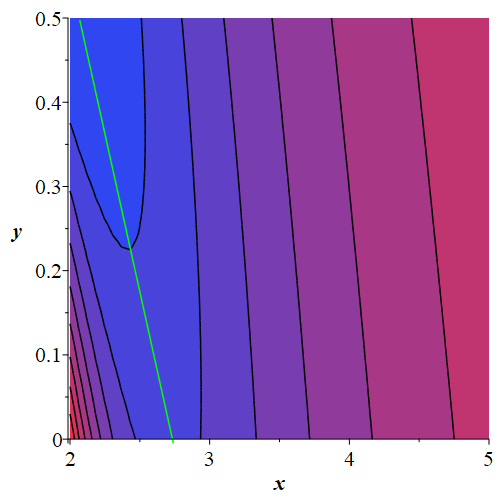}
\caption{Contour plot. `blue'$\rightarrow$`red' corresponds to `high'$\rightarrow$`low'.}\label{F:2b}
\end{subfigure}
\caption{The qualitative features of the spin zero Regge--Wheeler potential for the dominant multipole number $\ell=0$ are depicted.}
\label{F:2}
\end{figure}
%=====================================================
   
    The most notable feature is the spin zero peak; we see a slight shift in the peak locations between the spin one and spin zero potentials. The green line in Fig.~\ref{F:2b} is a `line of best fit', obtained \emph{via} manual corrections starting from the approximate location of the photon sphere as found in reference~\cite{Berry:2020}, and marks the $a$-dependent coordinate location $r_0\approx \frac{41}{15}m-\frac{4}{3}a$. Given one does not have information concerning how the peak shifts when comparing the spin one and spin zero potentials \emph{a priori}, and the peak location is not analytically solvable (see \S~\ref{S:QNMs}), this approximation is the best one can do in order to retain the desired level of mathematical tractability. Consequently, in \S~\ref{S:QNMs}, the approximation for $r_0$ as above is used in the computation of the relevant QNM profiles. The remaining features of the plot are qualitatively similar to those for the spin one case.
    \item Spin two bivector field (axial mode): The potential becomes
    \begin{equation}
        \V_2 = \left(1-\frac{2m\,\e^{-a/r}}{r}\right)\left\lbrace \frac{\ell(\ell+1)}{r^2}-\frac{2m\,\e^{-a/r}}{r^3}\left(3-\frac{a}{r}\right)\right\rbrace \ ,
    \end{equation}
    and, fixing the dominant multipole number $\ell=2$, one finds:
    \begin{equation}
        \eval{\V_2}_{\ell=2} = \frac{1}{r^2}\left(1-\frac{2m\,\e^{-a/r}}{r}\right)\left\lbrace 6-\frac{2m\,\e^{-a/r}}{r}\left(3-\frac{a}{r}\right)\right\rbrace \ .
    \end{equation}
    Once again, it is informative to re-express this in terms of the dimensionless variables $x=r/m$, $y=a/m$, giving
    \begin{equation}
        \eval{m^2\V_2}_{\ell=2} = \frac{1}{x^2}\left(1-\frac{2\,\e^{-y/x}}{x}\right)\left\lbrace 6 -\frac{2\,\e^{-y/x}}{x}\left(3-\frac{y}{x}\right)\right\rbrace \ .
    \end{equation}
    The qualitative features of $\V_2$ are then displayed in Fig.~\ref{F:3}.
    %
    %=====================================================
\begin{figure}[htb!]
%\vspace{-1cm}
\begin{subfigure}{.45\textwidth}
\includegraphics[scale=0.32]{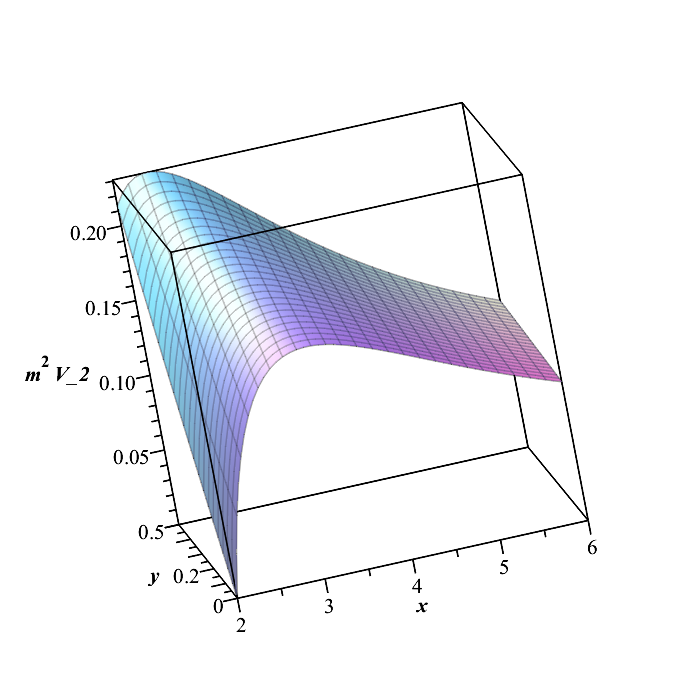}
\caption{Three-dimensional plot of $m^2\V_2$.}
\end{subfigure}
\begin{subfigure}{.45\textwidth}
\includegraphics[scale=0.36]{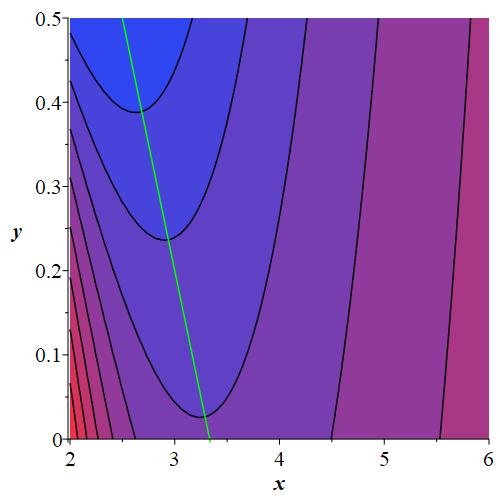}
\caption{Contour plot. `blue'$\rightarrow$`red' corresponds to `high'$\rightarrow$`low'.}\label{F:3b}
\end{subfigure}
\caption{The qualitative features of the spin two axial Regge--Wheeler potential for the dominant multipole number $\ell=2$ are depicted.}
\label{F:3}
\end{figure}
%=====================================================

    Once again the approximate location for the peak of the spin two (axial) potential is obtained \emph{via} application of manual corrections to the approximate location of the photon sphere as obtained in reference~\cite{Berry:2020}, and is found to be $r_2\approx\frac{10}{3}m-\frac{5}{3}a$ (this is the green line in Fig.~\ref{F:3b}). This approximation would serve as a starting point to extract QNM profile approximations for the spin two axial mode, similarly to the processes performed for spins one and zero in \S~\ref{S:QNMs}. However, for a combination of readability and tractability, this is for now a topic for future research. The remaining qualitative features of the spin two (axial) potential are similar to those for spins one and zero.
\end{itemize}

%========================================================
\section{First-order WKB approximation of the quasi-normal modes}\label{S:QNMs}
%========================================================

In order to calculate the quasi-normal modes for the candidate spacetime, one first defines them in the standard way: they are the \( \omega \) present in the right-hand-side of Eq.~\eqref{sepwaveform}, and they satisfy the ``radiation'' boundary conditions that \( \Psi \) is purely outgoing at spatial infinity and purely ingoing at the horizon \cite{Blome:1984, Santos:2019}.
Due to the inherent difficulty of analytically solving the Regge--Wheeler equation, a standard approach in the literature is to use the WKB approximation.
Although the WKB method was originally constructed to solve Schr\"odinger-type equations in quantum mechanics, the close resemblance between the Regge--Wheeler equation Eq.~\eqref{RWeq} and the Schr\"odinger equation allows for it to be readily adapted to the general relativistic setting. Given the use of the WKB approximation, one cannot extend the analysis of the QNMs for the candidate spacetime to the case when $a>2m/\e$, as for this case there are no horizons in the geometry. The existence of the outer horizon (or at the very least an extremal horizon) is critical to setting up the correct radiative boundary conditions. Other techniques for approximating the QNMs, \emph{e.g.} time domain integration (see~\cite{Churilova:2019} for an example), are likely to be applicable in this context. For now, this research is relegated to the domain of the future.

In order to proceed with the WKB method, one makes the stationary ansatz $\Psi\sim \e^{i\omega t}$, such that all of the qualitative behaviour for $\Psi$ is encoded in the profiles of the respective $\omega$. Computing a WKB approximation to first-order yields a relatively simple and tractable approximation to the quasi-normal modes for a black hole spacetime~\cite{Blome:1984, Santos:2019, Cuadros:2020}:
\begin{equation}
\omega^2 \approx \bigg[ \V(r_*) - i \left(n+\frac{1}{2}\right) \sqrt{-2 \, \partial_{r_*}^2 \V(r_*)} \bigg]_{r_*=r_{max}},
\end{equation}
where \( n \in \mathbb{N} \) is the overtone number, and where \( r_* = r_{max} \) is the tortoise coordinate location which maximises the relevant Regge--Wheeler potential. It is worth noting that $\eval{\V(r_*)}_{r_*=r_{max}}=\eval{\V(r)}_{r=r_{max}}$; this will be used in the subsequent analysis.

In-depth calculations of the WKB approximation up to higher orders in a general setting can also be found in references~\cite{Blome:1984, Santos:2019, Cuadros:2020}. Furthermore, various improvements and refinements to the WKB approximation have been explored in~\cite{Schutz:1985, Iyer:1986, Konoplya:2003, Zhidenko:2003, Matyjasek:2017, Konoplya:2019}. These include the derivation of a generalised higher-order formula, and the exploration of an improved variant of the original WKB approximation using Pad\'{e} approximants. These formulae present various intractabilities by hand and are best handled by numerical code. Consequently, a first-order calculation is performed to holistically analyse the qualitative aspects of the QNMs; refinement in accuracy is left to the numerical relativity community to explore further.

\subsection{Spin one}

For spin one particles recall that the relevant Regge--Wheeler potential is given by
\begin{equation}
    \V_1(r) = \left(1-\frac{2m\,\e^{-a/r}}{r}\right)\frac{\ell(\ell+1)}{r^2} \ .
\end{equation}
The $\V_1(r)$ Regge--Wheeler potential is proportional to the $V_0(r)$ effective potential used for determining the location of the photon sphere for massless particles~\cite{Berry:2020}. Specifically, one has
\begin{equation}
\pdv{\, \V_1}{r} = \frac{2\ell(\ell+1)}{r^3}\bigg\{ \frac{m \, \e^{-a/r}}{r}\left(3-\frac{a}{r}\right) - 1 \bigg\} \ .
\end{equation}
The resulting stationary points are not analytically solvable, and \emph{via} comparison with reference~\cite{Berry:2020} one sees that the spin one Regge--Wheeler potential is maximised precisely at the location of the photon sphere; \( r_1 = r_\gamma = \sqrt{ m\,\e^{-a/r_\gamma}(3r_\gamma-a) } \approx 3m - \frac{4}{3}a \).
Thus, one immediately obtains the spin one, first-order WKB approximation for the real part of the quasi-normal modes (recall $y=a/m$):
\begin{eqnarray}
\Re(\omega^2) \approx \eval{\V_1}_{r=r_1} &\approx& \frac{9 \, \ell(\ell+1)}{(9m-4a)^2} \left\lbrace 1 - \frac{6m\,\e^{\frac{3a}{4a-9m}}}{9m-4a} \right\rbrace \nonumber \\
&& \nonumber \\
&=& \frac{9\ell(\ell+1)}{m^2}\left\lbrace\frac{1}{(9-4y)^2}\left[1-\frac{6\,\e^{\frac{3y}{4y-9}}}{9-4y}\right]\right\rbrace \ .
\end{eqnarray}
Letting $r_c=3m-\frac{4}{3}a$, recalling $x=r/m$ (\emph{i.e.} $x_c=r_c/m$), and defining $z=a/r_c$, alternative representations include (eliminating $m$):
\begin{equation}
    \Re(\omega^2) \approx \frac{\ell(\ell+1)}{r_c^2}\left\lbrace 1 - \frac{2\,\e^{-a/r_c}(r_c+\frac{4}{3}a)}{3r_c}\right\rbrace = \frac{\ell(\ell+1)}{r_c^2}\left\lbrace 1 - \frac{2\,\e^{-z}(1+\frac{4}{3}z)}{3}\right\rbrace \ ,
\end{equation}
or (eliminating $a$):
\begin{equation}
    \Re(\omega^2) \approx \frac{\ell(\ell+1)}{r_c^2}\left\lbrace 1 - \frac{2m\,\e^{\frac{3(r_c-3m)}{4r_c}}}{r_c}\right\rbrace = \frac{\ell(\ell+1)}{r_c^2}\left\lbrace 1 - \frac{2\,\e^{\frac{3}{4}\left(1-\frac{3}{x_c}\right)}}{x_c}\right\rbrace \ .
\end{equation}
Which expression is preferred is a matter of taste and context.

Now, to compute the imaginary part of the QNMs, note that
\begin{equation}
\eval{ \pdv[2]{\V_1}{r_*} }_{r=r_1} = \eval{ \left(1-\frac{2m\,\e^{-a/r}}{r}\right) \left\lbrace\left(1-\frac{2m\,\e^{-a/r}}{r}\right)\pdv[2]{\V_1}{r} + \frac{2m\,\e^{-a/r}(r-a)}{r^3} \pdv{\V_1}{r} \right\rbrace }_{r=r_1} \ .
\end{equation}
But already it is known that \( \partial\V_1/\partial r \eval_{r=r_1} = 0 \), and so this reduces to
\begin{equation}
\eval{ \pdv[2]{\V_1}{r_*} }_{r=r_1} = \eval{ \left(1-\frac{2m\,\e^{-a/r}}{r}\right)^2 \pdv[2]{\V_1}{r} }_{r=r_1} \ .
\end{equation}
Thus, for spin one particles one finds
\begin{align}
\eval{ \pdv[2]{\V_1}{r_*} }_{r=r_1} &= \eval{ \ell(\ell+1) \left(1-\frac{2m\,\e^{-a/r}}{r}\right)^2 \left\{ \frac{6}{r^4} - (a-2r)(a-6r)\frac{2m\,\e^{-a/r}}{r^7} \right\} }_{r=r_1}	\notag \\
& \notag \\
&\approx \frac{486 \, \ell(\ell+1)}{(9m-4a)^4} \left(1 - \frac{6m\,\e^{3a\over4a-9m}}{9m-4a} \right)^2 	\notag \\
&\hspace{1.5cm}\times\left\{ 1-\frac{27m\,\e^{3a\over4a-9m} (18m-11a)(2m-a)}{(9m-4a)^3} \right\} \ ,
\end{align}
thereby giving
\begin{align}
\Im(\omega^2) &\approx \eval{ -\left(n+\frac{1}{2}\right)\sqrt{-2 \, \partial_{r_*}^2 \V_1(r)} }_{r=r_1} 	\notag \\
& \notag \\
&\approx -\frac{18\sqrt{3}}{(9m-4a)^2} \, \left(n+\frac{1}{2}\right) \left(1 - \frac{6m\,\e^{3a\over4a-9m}}{9m-4a} \right)  	\notag \\
&\hspace{1cm}\times\sqrt{\ell(\ell+1) \left\lbrace \frac{27m\,\e^{3a\over4a-9m} (18m-11a)(2m-a)}{(9m-4a)^3} - 1 \right\rbrace} \notag \\
& \notag \\
&= -\frac{\left(n+\frac{1}{2}\right)}{m^2}18\sqrt{3}\left(1-\frac{6\,\e^{\frac{3y}{4y-9}}}{9-4y}\right) \notag \\
&\hspace{1cm}\times\sqrt{\frac{\ell(\ell+1)}{(9-4y)^4}\left\lbrace\frac{27\,\e^{\frac{3y}{4y-9}}(18-11y)(2-y)}{(9-4y)^3}-1\right\rbrace} \ .
\end{align}
Alternative expressions include (eliminating $m$):
\begin{eqnarray}
    \Im(\omega^2) &\approx& -\frac{2\sqrt{3}\left(n+\frac{1}{2}\right)}{r_c^2}\left(1-\frac{2\,\e^{-a/r_c}(3r_c+4a)}{9r_c}\right) \nonumber \\
    && \nonumber \\
    && \hspace{1cm}\times \sqrt{\ell(\ell+1)\left\lbrace\frac{(3r_c+4a)(2r_c-a)(6r_c-a)\,\e^{-a/r_c}}{27r_c^3}-1\right\rbrace} \nonumber \\
    && \nonumber \\
    &=& -\frac{2\sqrt{3}\left(n+\frac{1}{2}\right)}{r_c^2}\left(1-\frac{2\,\e^{-z}(3+4z)}{9}\right) \nonumber \\
    && \nonumber \\
    &&\hspace{1cm}\times\sqrt{\ell(\ell+1)\left\lbrace\frac{(3+4z)(2-z)(6-z)\,\e^{-z}}{27}-1\right\rbrace} \ ,
\end{eqnarray}
or (eliminating $a$):
\begin{eqnarray}
    \Im(\omega^2) &\approx& -\frac{2\sqrt{3}\left(n+\frac{1}{2}\right)}{r_c^2}\left(1-\frac{2m\,\e^{\frac{3(r_c-3m)}{4r_c}}}{r_c}\right) \nonumber \\
    && \nonumber \\
    && \hspace{1cm}\times \sqrt{\ell(\ell+1)\left\lbrace\frac{3m\,\e^{\frac{3(r_c-3m)}{4r_c}}(11r_c-9m)(3r_c-m)}{16r_c^3}-1\right\rbrace} \nonumber \\
    && \nonumber \\
    &=& -\frac{2\sqrt{3}\left(n+\frac{1}{2}\right)}{r_c^2}\left(1-\frac{2\,\e^{\frac{3}{4}\left(1-\frac{3}{x_c}\right)}}{x_c}\right) \nonumber \\
    && \nonumber \\
    &&\hspace{1cm}\times\sqrt{\ell(\ell+1)\left\lbrace\frac{3\,\e^{\frac{3}{4}\left(1-\frac{3}{x_c}\right)}\left(11x_c-9\right)\left(3x_c-1\right)}{16x_c^3} - 1 \right\rbrace} \ .
\end{eqnarray}
Thus, the first-order WKB approximation of the spin one QNMs, for general multipole numbers $\ell$ and oscillation modes $n$, is given by
\begin{eqnarray}\label{spinoneapprox}
    \omega^2 &\approx& \frac{9\ell(\ell+1)}{(9m-4a)^2}\left(1-\frac{6m\,\e^{\frac{3a}{4a-9m}}}{9m-4a}\right) \nonumber \\
    && \nonumber \\
    &&\times\left\lbrace 1 - 2\sqrt{3}i\left(n+\frac{1}{2}\right)\sqrt{\frac{1}{\ell(\ell+1)}\left[\frac{27m\,\e^{\frac{3a}{4a-9m}}(18m-11a)(2m-a)}{(9m-4a)^3}-1\right]}\right\rbrace \nonumber \\
    && \nonumber \\
    &=& \frac{9\ell(\ell+1)}{m^2(9-4y)^2}\left(1-\frac{6\,\e^{\frac{3y}{4y-9}}}{9-4y}\right) \nonumber \\
    && \nonumber \\
    &&\times\left\lbrace 1 - 2\sqrt{3}i\left(n+\frac{1}{2}\right)\sqrt{\frac{1}{\ell(\ell+1)}\left[\frac{27\,\e^{\frac{3y}{4y-9}}(18-11y)(2-y)}{(9-4y)^3}-1\right]}\right\rbrace \ . \nonumber \\
    &&
\end{eqnarray}
For the purposes of extracting numerical results in \S~\ref{S:Results}, it is useful to redefine this as the dimensionless quantity
\begin{eqnarray}\label{eqforspinoneres}
    m^2\omega^2 &\approx& \frac{9\ell(\ell+1)}{(9-4y)^2}\left(1-\frac{6\,\e^{\frac{3y}{4y-9}}}{9-4y}\right) \nonumber \\
    && \nonumber \\
    &&\times\left\lbrace 1 - 2\sqrt{3}i\left(n+\frac{1}{2}\right)\sqrt{\frac{1}{\ell(\ell+1)}\left[\frac{27\,\e^{\frac{3y}{4y-9}}(18-11y)(2-y)}{(9-4y)^3}-1\right]}\right\rbrace \ . \nonumber \\
    &&
\end{eqnarray}
Alternative expressions include (eliminating $m$):
\begin{eqnarray}
    \omega^2 &\approx& \frac{\ell(\ell+1)}{r_c^2}\left(1-\frac{2(3r_c+4a)\,\e^{-a/r_c}}{9r_c}\right) \nonumber \\
    && \nonumber \\
    && \times \left\lbrace 1-2\sqrt{3}i\left(n+\frac{1}{2}\right)\sqrt{\frac{1}{\ell(\ell+1)}\left[\frac{(2r_c-a)(6r_c-a)(3r_c+4a)\,\e^{-a/r_c}}{27r_c^3}-1\right]}\right\rbrace \nonumber \\
    && \nonumber \\
    &=& \frac{\ell(\ell+1)}{r_c^2}\left(1-\frac{2(3+4z)\,\e^{-z}}{9}\right) \nonumber \\
    && \nonumber \\
    &&\times\left\lbrace 1 - 2\sqrt{3}i\left(n+\frac{1}{2}\right)\sqrt{\frac{1}{\ell(\ell+1)}\left[\frac{(3+4z)(2-z)(6-z)\e^{-z}}{27}-1\right]}\right\rbrace \ , \nonumber \\
    &&
\end{eqnarray}
or (eliminating $a$):
\begin{eqnarray}
    \omega^2 &\approx& \frac{\ell(\ell+1)}{r_c^2}\left(1-\frac{2m\,\e^{\frac{3(r_c-3m)}{4r_c}}}{r_c}\right) \nonumber \\
    && \nonumber \\
    && \times \left\lbrace 1-2\sqrt{3}i\left(n+\frac{1}{2}\right)\sqrt{\frac{1}{\ell(\ell+1)}\left[\frac{3m\,\e^{\frac{3(r_c-3m)}{4r_c}}(11r_c-9m)(3r_c-m)}{16r_c^3}-1\right]}\right\rbrace \nonumber \\
    && \nonumber \\
    &=& \frac{\ell(\ell+1)}{r_c^2}\left(1-\frac{2\,\e^{\frac{3}{4}\left(1-\frac{3}{x_c}\right)}}{x_c}\right) \nonumber \\
    && \nonumber \\
    &&\times\left\lbrace 1 - 2\sqrt{3}i\left(n+\frac{1}{2}\right)\sqrt{\frac{1}{\ell(\ell+1)}\left[\frac{3\,\e^{\frac{3}{4}\left(1-\frac{3}{x_c}\right)}(11x_c-9)(3x_c-1)}{16x_c^3}-1\right]}\right\rbrace \ . \nonumber \\
    &&
\end{eqnarray}
In the Schwarzschild limit one obtains
\begin{equation}
\omega^2_{Sch.} = \lim_{a\rightarrow0} \left(\omega^2\right) \approx \frac{\ell(\ell+1)}{27m^2}\left( 1 - \frac{i(2n+1)}{\sqrt{\ell(\ell+1)}} \right),
\end{equation}
which agrees with existing work in the literature \cite{Blome:1984, Santos:2019}.

%%%%%%

\subsection{Spin zero}

For spin zero particles recall that one has the following specific form for the Regge--Wheeler potential:
\begin{equation}
    \mathcal{V}_{0}(r) = \left(1-\frac{2m\,\e^{-a/r}}{r}\right)\left\lbrace\frac{\ell(\ell+1)}{r^{2}}+\frac{2m\,\e^{-a/r}\left(1-\frac{a}{r}\right)}{r^{3}}\right\rbrace \ .
\end{equation}
It is immediately clear that the peak of this potential is going to be slightly shifted from the location of the photon sphere, which for the spin one case maximised $\mathcal{V}_{1}(r)$. Computing:
\begin{eqnarray}
    \frac{\partial{\mathcal{V}_{0}}}{\partial{r}} &=& \frac{1}{r^3}\Bigg\lbrace\left[\frac{2m\,\e^{-a/r}}{r}\left(3-\frac{a}{r}\right)-2\right]\left[\ell(\ell+1)+\frac{2m\,\e^{-a/r}}{r}\left(1-\frac{a}{r}\right)\right] \nonumber \\
    && \nonumber \\
    && + \left(1-\frac{2m\,\e^{-a/r}}{r}\right)\left[\frac{-2m\,\e^{-a/r}}{r}\left(\left(\frac{a}{r}\right)^2-3\left(\frac{a}{r}\right)+1\right)\right]\Bigg\rbrace \ .
\end{eqnarray}
The associated stationary points are not analytically solvable for $r$. It is worth noting that in general, the stationary points of $\V_0$ are $\ell$-dependent, unlike in the case for $\V_1$. Without knowledge of the location of the peak for the spin zero potential \emph{a priori}, the best line of inquiry which retains the desired level of tractability is to specialise to the scalar $s$-wave (corresponding to $\ell=0$), which is of particular importance, playing a dominant role in the signal. This constraint is fit for purpose in extracting the relevant results in \S~\ref{S:Results}. Specialising to the $s$-wave, one finds
\begin{equation}
    \eval{\frac{\partial\V_0}{\partial r}}_{\ell=0} = \frac{2m\,\e^{-a/r}}{r^4}\left\lbrace\frac{m\,\e^{-a/r}}{r}\left[4\left(\frac{a}{r}\right)^2-14\left(\frac{a}{r}\right)+8\right]-\left[\left(\frac{a}{r}\right)^2-5\left(\frac{a}{r}\right)+3\right]\right\rbrace \ ,
\end{equation}
and still the associated stationary points are not analytically solvable. As such, to make progress one uses the approximate location of the peak as found in \S~\ref{S:RW_potential}; $r_0\approx\frac{41}{15}m-\frac{4}{3}a$. See Fig.~\ref{F:2b} for details. One obtains the following approximation for the real part of the spin zero QNMs for the scalar $s$-wave:
\begin{eqnarray}\label{Respin0}
    \Re(\omega^2) \approx \eval{\V_0(r_0)}_{\ell=0} &\approx& \frac{6750m\,\e^{\frac{15a}{20a-41m}}(41m-35a)\left(30m\,\e^{\frac{15a}{20a-41m}}+20a-41m\right)}{(20a-41m)^5} \nonumber \\
    && \nonumber \\
    &=& \frac{6750}{m^2}\left[\frac{\e^{\frac{15y}{20y-41}}(41-35y)(30\,\e^{\frac{15y}{20y-41}}+20y-41)}{(20y-41)^5}\right] \ .
\end{eqnarray}
For the imaginary part of the spin zero QNMs, firstly one has
\begin{eqnarray}
    \eval{\frac{\partial^2\V_0}{\partial r_*^2}}_{r=r_0} = \eval{\left(1-\frac{2m\,\e^{-a/r}}{r}\right)^2\frac{\partial^2\V_0}{\partial r^2}}_{r=r_0} \ ,
\end{eqnarray}
with
\begin{eqnarray}
    \frac{\partial^2\V_0}{\partial r^2} &=& \frac{2}{r^4}\Bigg\lbrace 3\ell(\ell+1) + \frac{4m^2\,\e^{-2a/r}}{r^2}\left[2\left(\frac{a}{r}\right)^2-10\left(\frac{a}{r}\right)+5\right]\left(\frac{a}{r}-2\right) \nonumber \\
    && \nonumber \\
    && -\frac{m\,\e^{-a/r}}{r}\left[12(\ell^2+\ell-1)-4\left(\frac{a}{r}\right)(2\ell^2+2\ell-7)+\left(\frac{a}{r}\right)^2(\ell^2+\ell-11)+\left(\frac{a}{r}\right)^3\right]\Bigg\rbrace \ , \nonumber \\
    &&
\end{eqnarray}
giving
\begin{eqnarray}
    \frac{\partial^2\V_0}{\partial r_*^2} &=& \frac{2}{r^4}\left(1-\frac{2m\,\e^{-a/r}}{r}\right)^2\Bigg\lbrace 3\ell(\ell+1) + \frac{4m^2\,\e^{-2a/r}}{r^2}\left[2\left(\frac{a}{r}\right)^2-10\left(\frac{a}{r}\right)+5\right]\left(\frac{a}{r}-2\right) \nonumber \\
    && \nonumber \\
    && -\frac{m\,\e^{-a/r}}{r}\left[12(\ell^2+\ell-1)-4\left(\frac{a}{r}\right)(2\ell^2+2\ell-7)+\left(\frac{a}{r}\right)^2(\ell^2+\ell-11)+\left(\frac{a}{r}\right)^3\right]\Bigg\rbrace \ . \nonumber \\
    &&
\end{eqnarray}
For tractability, it is now prudent to specialise to $\ell=0$. This yields
\begin{eqnarray}
    \eval{ \frac{\partial^2\V_0}{\partial r_*^2}}_{\ell=0} &=& \frac{2}{r^4}\left(1-\frac{2m\,\e^{-a/r}}{r}\right)^2\Bigg\lbrace\left(\frac{2m\,\e^{-a/r}}{r}\right)^2\left(\frac{a}{r}-2\right)\left[2\left(\frac{a}{r}\right)^2-10\left(\frac{a}{r}\right)+5\right] \nonumber \\
    && \nonumber \\
    && - \frac{m\,\e^{-a/r}}{r}\left[-12+28\left(\frac{a}{r}\right)-11\left(\frac{a}{r}\right)^2+\left(\frac{a}{r}\right)^3\right]\Bigg\rbrace \ ,
\end{eqnarray}
and one obtains the following approximation for the imaginary part of the $s$-wave spin zero QNMs (expressed only as a function of $y$ for readability):
\begin{eqnarray}\label{Imspin0}
    \Im(\omega^2) &\approx& \eval{ -\left(n+\frac{1}{2}\right)\sqrt{-2\,\left[\partial_{r_*}^{2}\V_0\left(r_0\approx\frac{41}{15}m-\frac{4}{3}a\right)\right]}}_{\ell=0} \nonumber \\
    && \nonumber \\
    &=& -\frac{1350}{m^2}\left(n+\frac{1}{2}\right)\left[\frac{\sqrt{5}\,\e^{\frac{15y}{20y-41}}(30\,\e^{\frac{15y}{20y-41}}+20y-41)}{(20y-41)^{\frac{11}{2}}}\right] \nonumber \\
    && \nonumber \\
    &&\times\Bigg\lbrace 100\,\e^{\frac{15y}{20y-41}}(59950y^3-247230y^2+327795y-137842) \nonumber \\
    && \nonumber \\
    && +2112500y^4-13535125y^3+31644825y^2-31703660y+11303044\Bigg\rbrace^{\frac{1}{2}} \ . \nonumber \\
    &&
\end{eqnarray}
Combining Eq.~(\ref{Respin0}) and Eq.~(\ref{Imspin0}) gives the approximation for the dimensionful $\omega^2$, however the expression is unwieldy and not particularly important to display here. Using Eq's.~(\ref{Respin0}) and (\ref{Imspin0}) to compute the real and imaginary approximations for the dimensionless $m^2\omega^2$ respectively is sufficient for extracting the relevant results in \S~\ref{S:Results}.

%=============================================
\section{Numerical results}\label{S:Results}
%=============================================

Given that the behaviour of the waveform is aggressively governed by the fundamental mode, to extract profile approximations it is both physically well-motivated and mathematically tractable to specialise to $n=0$. Given the WKB approximation is only valid in the presence of a nontrivial horizon structure, and that the candidate spacetime only possesses horizons for $a\in\left(0,\frac{2m}{\e}\right)$, it is prudent for one to define the dimensionless object $\hat{a}=a/a_{max}=\frac{a\e}{2m}=\frac{\e}{2}y$, such that $\hat{a}\in(0,1)$. One may then define the dimensionless $\hat{\omega}=\omega m$, such that all of the qualitative information for the dimensionful $\omega$ as a function of the dimensionful $a$ is now encoded in the dimensionless $\hat{\omega}$ as a function of the dimensionless $\hat{a}$. As such, one then examines $\hat{\omega}$ by plugging in values of $y=\frac{2}{\e}\hat{a}$ into the relevant equations from \S~\ref{S:QNMs} on a case-by-case basis. Lastly, it will be of most use to analyse the dominant multipole number in each spin case. Given $\ell\geq S$, for electromagnetic spin one fluctuations this will correspond to analysing $\ell=1$, for scalar spin zero fluctuations one analyses the $s$-wave corresponding to fixing $\ell=0$ (as already stipulated), and finally for spin two axial perturbations one would fix $\ell=2$. Notably this implies that the approximate locations for the peaks of the relevant RW-potentials computed in \S~\ref{S:RW_potential} are directly applicable here.

\subsection{Spin one}

Consequently, to analyse QNM profile approximations for electromagnetic spin one fluctuations on the background spacetime, fix the fundamental mode $n=0$, and analyse the special case of the dominant multipole number $\ell=1$. Substituting these values into Eq.~(\ref{eqforspinoneres}) and computing the resulting square root gives the results from Table~\ref{tab:1} for the approximation of $\hat{\omega}$ for different values of $\hat{a}\in(0,1)$ (rounded to $6$ d.p.):

%=============
\begin{table}[!ht]
\begin{center}
\caption{Fundamental QNM of the spin one field for $\ell=1$, obtained \emph{via} first-order WKB approximation.}\label{tab:1}
\hspace{-35pt}
\begin{tabular}{||c|c||}
\hline
\hline
\vphantom{\bigg|} $\hat{a}$ & WKB approx. for $\hat{\omega}$ \\
\hline
\hline
$0.0$ & $0.287050-0.091235i$ \\
\hline
$0.1$ & $0.293902-0.092012i$ \\
\hline
$0.2$ & $0.301291-0.092532i$ \\
\hline
$0.3$ & $0.309304-0.092708i$ \\
\hline
$0.4$ & $0.318051-0.092419i$ \\
\hline
$0.5$ & $0.327658-0.091486i$ \\
\hline
$0.6$ & $0.338285-0.089636i$ \\
\hline
$0.7$ & $0.350117-0.086433i$ \\
\hline
$0.8$ & $0.363377-0.081139i$ \\
\hline
$0.9$ & $0.378330-0.072338i$ \\
\hline
$1.0$ & $0.395289-0.056624i$ \\
\hline
\hline
\end{tabular}
\end{center}
\end{table}
%================
%================
\begin{figure}[htb!]
\begin{center}
%\vspace{-1cm}
\includegraphics[scale=0.35]{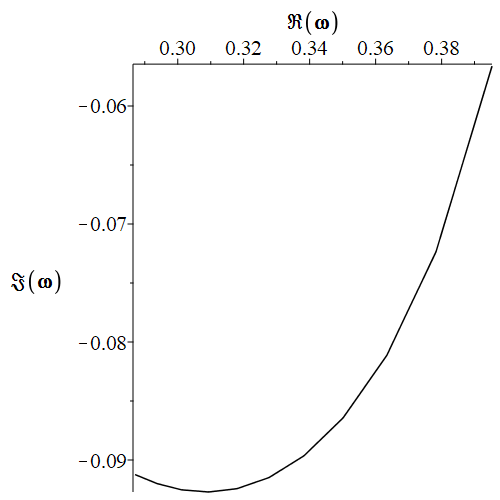}
\caption{A plot of the points from Table~\ref{tab:1} with a linear interpolation curve. $\mbox{Re}(\hat{\omega})$ increases monotonically with $\hat{a}$, hence increasing $\hat{a}$ corresponds to one moving from left to right.}\label{F:4}
\end{center}
\end{figure}
%=====================================================

Immediately there are the following qualitative observations:
\begin{itemize}
    \item As a sanity check, $\Im(\hat{\omega})<0$ for all values of $\hat{a}$, indicating that the propagation of electromagnetic fields in the background spacetime is stable --- an expected result;
    \item $\Re(\hat{\omega})$ increases monotonically with $\hat{a}$ --- this is the frequency of the corresponding QNMs;
    \item $\Im(\hat{\omega})$ decreases with $\hat{a}$ initially, up until $\hat{a}\approx 0.35$, and then it increases monotonically with $\hat{a}$ for the remainder of the domain --- this is the decay rate or damping rate of the QNMs;
    \item Given that throughout the historical literature, it is conventional to assert $\hat{a}\sim m_{p}$, it is likely of primary interest to examine the behaviour of this plot for small $\hat{a}$. In view of this, if one constrains the analysis of the qualitative behaviour for $\hat{\omega}$ \emph{prior} to the trough present in Fig.~\ref{F:4}, one would expect that the signals for electromagnetic radiation propagating in the presence of a regular black hole with asymptotically Minkowski core should have both a higher frequency as well as a faster decay rate than their Schwarzschild counterparts. This qualitative result \emph{may} translate to the spin two case, and speak directly to the LIGO/VIRGO calculation. The fact that the signal is expected to be shorter-lived could present a heightened level of experimental difficulty when trying to delineate signals, though this may very well be offset by the fact that the signal also carries higher energy; further discussion on these points is left to both the numerical relativity and experimental communities.
\end{itemize}

%%%%%%

\subsection{Spin zero}

For spin zero scalar fluctuations, specialising to the $s$-wave, similarly fix the fundamental mode $n=0$. Substituting this into Eq.~(\ref{Respin0}) and Eq.~(\ref{Imspin0}), which are the relevant equations to compute the real and imaginary approximations of $\hat{\omega}^2$ respectively (recall these have already specialised to the $s$-wave given $\ell$ has already been fixed to be zero), and then taking the appropriate square root yields the results from Table~\ref{tab:2} (to $6$ d.p.):

%=============
\begin{table}[!ht]
\begin{center}
\caption{Fundamental QNM of the massless, minimally coupled spin zero scalar field for the $s$-wave ($\ell=0$), obtained \emph{via} first-order WKB approximation.}\label{tab:2}
\hspace{-35pt}
\begin{tabular}{||c|c||}
\hline
\hline
\vphantom{\bigg|} $\hat{a}$ & WKB approx. for $\hat{\omega}$ \\
\hline
\hline
$0.0$ & $0.187409-0.094054i$\\
\hline
$0.1$ & $0.189734-0.094530i$ \\
\hline
$0.2$ & $0.191948-0.094669i$ \\
\hline
$0.3$ & $0.194049-0.094425i$ \\
\hline
$0.4$ & $0.196027-0.093742i$ \\
\hline
$0.5$ & $0.197868-0.092557i$ \\
\hline
$0.6$ & $0.199552-0.090796i$ \\
\hline
$0.7$ & $0.201042-0.088385i$ \\
\hline
$0.8$ & $0.202285-0.085306i$ \\
\hline
$0.9$ & $0.203235-0.081735i$ \\
\hline
$1.0$ & $0.203894-0.078421i$ \\
\hline
\hline
\end{tabular}
\end{center}
\end{table}
%================
%================
\begin{figure}[htb!]
\begin{center}
%\vspace{-1cm}
\includegraphics[scale=0.35]{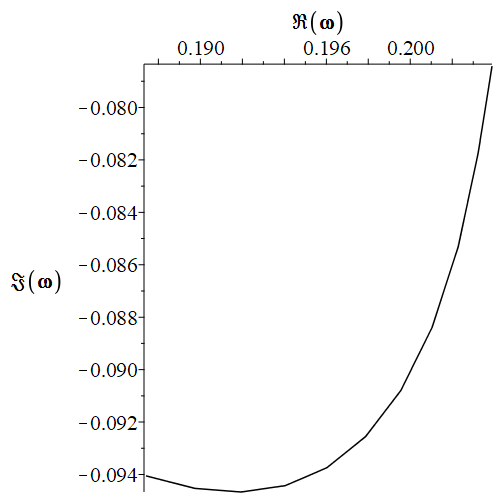}
\caption{A plot of the points from Table~\ref{tab:2} with a linear interpolation curve. $\mbox{Re}(\hat{\omega})$ increases monotonically with $\hat{a}$, hence increasing $\hat{a}$ corresponds to one moving from left to right.}\label{F:5}
\end{center}
\end{figure}
%=====================================================

There are the following qualitative observations:
\begin{itemize}
    \item $\Re(\hat{\omega})$ once again increases monotonically with $\hat{a}$ --- higher $\hat{a}$-values correspond to higher frequency fundamental modes;
    \item $\Im(\hat{\omega})<0$ for all $\hat{a}$, indicating that the $s$-wave for minimally coupled massless scalar fields propagating in the background spacetime is stable;
    \item $\Im(\hat{\omega})$ decreases with $\hat{a}$ initially (down to a trough around $\hat{a}=0.25$), before monotonically increasing with $\hat{a}$ for the rest of the domain --- this is the decay/damping rate of the QNMs;
    \item Similarly as for the electromagnetic spin one case, when one examines the behaviour for small $\hat{a}$, the signals for the fundamental mode of spin zero scalar field perturbations in the presence of a regular black hole with asymptotically Minkowski core is expected to have a higher frequency and to be shorter-lived than for their Schwarzschild counterparts.
\end{itemize}

%========================================================
\section{Perturbing the potential --- general first-order analysis}\label{S:RWperturb}
%========================================================

Suppose one perturbs the Regge--Wheeler potential itself, replacing $\V(r) \to \V(r) + \delta\V(r)$. It is of interest to analyse what effect this has on the estimate for the QNMs. Classical perturbation of the potential to first-order in $\epsilon$ is performed, capturing any linear contributions from external agents that may disturb the propagating waveforms. First-order perturbation is well-motivated from the perspective of the historical literature, and ensures the analysis has the desired level of tractability. As such, one has the following: $\V(r) \to \V(r) + \delta\V(r) = \V(r) + \epsilon\,\delta\V_a(r) + \epsilon^2\,\delta\V_b(r) + \mathcal{O}(\epsilon^3) \approx \V(r) + \epsilon\,\delta\V_a(r)$. All terms of order $\epsilon^2$ or higher are therefore truncated. Consequently, for notational convenience it is advantageous to simply replace $\delta\V_\alpha(r)$ with $\delta\V(r)$ in the discourse that follows, eliminating superfluous indices. Also for notational convenience, define $r_{max}=r_\sigma$ to be the generalised location of the peak of the potentials. One observes the following effects on the QNMs:

\begin{itemize}
\item 
Firstly, the \emph{position} of the peak shifts:
\begin{equation}
0 \approx [\V+\epsilon\,\delta\V]'(r)\Big|_{r=r_\sigma+\epsilon\,\delta r_\sigma} \ ,
\end{equation}
giving
\begin{equation}\label{perturb1}
\V'(r_\sigma+\epsilon\,\delta r_\sigma) + \epsilon\,[\delta\V]'(r_\sigma+\epsilon\,\delta r_\sigma) \approx 0 \ .
\end{equation}
Performing a first-order Taylor series expansion of the left-hand-side of Eq.~(\ref{perturb1}) about $\delta r_0=0$ then yields
\begin{equation}
    \V'(r_\sigma) + \epsilon\,[\delta\V]'(r_\sigma) + \epsilon\,\delta r_\sigma\left\lbrace \V''(r_\sigma) + \epsilon\,[\delta\V]''(r_\sigma)\right\rbrace \approx 0 \ ,
\end{equation}
and eliminating the term of order $\epsilon^2$, combined with the knowledge that $\V'(r_\sigma)=0$, gives
\begin{equation}\label{deltarsig}
    \delta r_\sigma \approx -\frac{[\delta\V]'(r_\sigma)}{\V''(r_\sigma)} \ .
\end{equation}
\item 
Secondly, the \emph{height} of the peak shifts:
\begin{equation}
[\V+\epsilon\,\delta\V](r_\sigma+\epsilon\,\delta r_\sigma)= \V(r_\sigma+\epsilon\,\delta r_\sigma) + \epsilon\,[\delta\V](r_\sigma+\epsilon\,\delta r_\sigma) \ ,
\end{equation}
and performing a first-order Taylor series expansion about $\delta r_0=0$ yields the following to first-order in $\epsilon$:
\begin{eqnarray}
[\V+\epsilon\,\delta\V](r_\sigma+\epsilon\,\delta r_\sigma) &\approx& \V(r_\sigma) + \epsilon\,\left\lbrace[\delta\V](r_\sigma)+\delta r_\sigma\V'(r_\sigma)\right\rbrace \nonumber \\
&& \nonumber \\
&=& \V(r_\sigma) + \epsilon\,[\delta\V](r_\sigma) \ .
\end{eqnarray}
\item
Third, the \emph{curvature} at the peak shifts
\begin{equation}
[\V+\epsilon\,\delta\V]''(r_\sigma+\epsilon\,\delta r_\sigma) = \V''(r_\sigma+\epsilon\,\delta r_\sigma) + \epsilon\,[\delta\V]''(r_\sigma+\epsilon\,\delta r_\sigma) \ ,
\end{equation}
which for first-order-Taylor about $\delta r_\sigma=0$ and to first-order in $\epsilon$ gives
\begin{equation}
    [\V+\epsilon\,\delta\V]''(r_\sigma+\epsilon\,\delta r_\sigma) \approx \V''(r_\sigma) + \epsilon\,[\delta\V]''(r_\sigma) + \epsilon\,\delta r_\sigma\V'''(r_\sigma) \ ,
\end{equation}
which from Eq.~(\ref{deltarsig}) can then be approximated by the following
\begin{equation}\label{Vprimeprime}
    [\V+\epsilon\,\delta\V]''(r_\sigma+\epsilon\,\delta r_\sigma) \approx \V''(r_\sigma) + \epsilon\,\left\lbrace[\delta\V]''(r_\sigma) - \frac{\V'''(r_\sigma)}{\V''(r_\sigma)}[\delta\V]'(r_\sigma)\right\rbrace \ .
\end{equation}
\end{itemize}

Given the following (one assumes the existence of some tortoise coordinate relation $\partial_{r_*}=T(r)\partial_{r}$):
\begin{eqnarray}
    \partial_{r_*}^2[\V+\epsilon\,\delta\V](r_\sigma+\epsilon\,\delta r_\sigma) &=& \eval{\partial_{r_*}\left\lbrace T(r)\partial_{r}[\V+\epsilon\,\delta\V](r)\right\rbrace}_{r=r_\sigma+\epsilon\,\delta r_\sigma} \nonumber \\
    && \nonumber \\
    &=& \eval{T^2(r)[\V+\epsilon\,\delta\V]''(r)+[\V+\epsilon\,\delta\V]'(r)T'(r)T(r)}_{r=r_\sigma+\epsilon\,\delta r_\sigma} \ , \nonumber \\
    &&
\end{eqnarray}
and performing Taylor series expansions about $\epsilon=0$ for the relevant terms and truncating to first-order in $\epsilon$ gives the following:
\begin{eqnarray}
    \partial_{r_*}^2[\V+\epsilon\,\delta\V](r_\sigma+\epsilon\,\delta r_\sigma) &=& [T(r_\sigma+\epsilon\,\delta r_\sigma)]^2[\V+\epsilon\,\delta\V]''(r_\sigma+\epsilon\,\delta r_\sigma) \nonumber \\
    && \nonumber \\
    && \qquad + [\V+\epsilon\,\delta\V]'(r_\sigma+\epsilon\,\delta r_\sigma)T'(r_\sigma+\epsilon\,\delta r_\sigma)T(r_\sigma+\epsilon\,\delta r_\sigma) \nonumber \\
    && \nonumber \\
    &\approx& \left[T(r_\sigma)+\epsilon\,\delta r_\sigma T'(r_\sigma)\right]^2\,[\V+\epsilon\,\delta\V]''(r_\sigma+\epsilon\,\delta r_\sigma) \nonumber \\
    && \nonumber \\
    && \qquad \qquad \qquad \qquad \qquad \qquad + \epsilon\,\delta r_\sigma\V''(r_\sigma)T'(r_\sigma)T(r_\sigma) \nonumber \\
    && \nonumber \\
    &\approx& [T^2(r_\sigma) + 2\epsilon\,\delta r_\sigma T'(r_\sigma)][\V+\epsilon\,\delta\V]''(r_\sigma+\epsilon\,\delta r_\sigma) \nonumber \\
    && \nonumber \\
    && \qquad \qquad \qquad \qquad \qquad \qquad - \epsilon\,T'(r_\sigma)T(r_\sigma)[\delta\V]'(r_\sigma) \ , \nonumber \\
    &&
\end{eqnarray}
and substituting the result from Eq.~(\ref{Vprimeprime}) then finally gives (all functions on the right-hand-side are evaluated at $r=r_\sigma$; notation suppressed for tractability):
\begin{eqnarray}
    \partial_{r_*}^2[\V+\epsilon\,\delta\V] \approx T^2\V'' - \epsilon\,\left\lbrace\left(T'T+2T'+T^2\frac{\V'''}{\V''}\right)[\delta\V]'-T^2[\delta\V]''\right\rbrace \ .
\end{eqnarray}
As such, for the square root one has:
\begin{eqnarray}
    \sqrt{-2\partial_{r_*}^2[\V+\epsilon\,\delta\V](r_\sigma+\epsilon\,\delta r_\sigma)} &\approx& \Bigg\lbrace-2T^2\V'' \nonumber \\
    && \nonumber \\
    && \ + 2\epsilon\,\left[\left(T'T+2T'+T^2\frac{\V'''}{\V''}\right)[\delta\V]'-T^2[\delta\V]''\right]\Bigg\rbrace^{\frac{1}{2}} \ , \nonumber \\
    &&
\end{eqnarray}
and performing a first-order Taylor series expansion about $\epsilon=0$ gives the following (all functions on the right-hand-side are evaluated at $r=r_\sigma$):
\begin{eqnarray}
    \sqrt{-2\partial_{r_*}^2[\V+\epsilon\,\delta\V]} &\approx& T\sqrt{-2\V''}-\frac{\epsilon}{\sqrt{2}}\Bigg\lbrace\left(\frac{T'}{\vert\V''\vert^{\frac{1}{2}}}+\frac{2T'}{T\vert\V''\vert^{\frac{1}{2}}}+\frac{T\V'''}{\vert\V''\vert^{\frac{3}{2}}}\right)[\delta\V]' \nonumber \\
    && \nonumber \\
    && \qquad \qquad \qquad \qquad \qquad \qquad \qquad \qquad -\frac{T}{\vert\V''\vert^{\frac{1}{2}}}[\delta\V]''\Bigg\rbrace \ , \nonumber \\
    &&
\end{eqnarray}
assembling the pieces, to first-order in WKB and first-order in $\epsilon$ the approximate shift in the QNMs is given by:
\begin{eqnarray}\label{perturbativeform}
    \delta(\omega^2_n) &\approx& \epsilon\,[\delta\V] \nonumber \\
    && \qquad + i\left(n+\frac{1}{2}\right)\frac{\epsilon}{\sqrt{2}}\left\lbrace\left(\frac{T'}{\vert\V''\vert^{\frac{1}{2}}}+\frac{2T'}{T\vert\V''\vert^{\frac{1}{2}}}+\frac{T\V'''}{\vert\V''\vert^{\frac{3}{2}}}\right)[\delta\V]'-\frac{T}{\vert\V''\vert^{\frac{1}{2}}}[\delta\V]''\right\rbrace \ , \nonumber \\
    &&
\end{eqnarray}
where all expressions on the right-hand-side are evaluated at $r=r_\sigma$. This specific formula is general for all instances where the WKB approximation is appropriate. It is informative to now apply this to the most straightforward example of Schwarzschild spacetime.

\paragraph{Perturbing around Schwarzschild:}\

For the particular case of spin one Schwarzschild, one sets $a\rightarrow0$, and has the following:
\begin{equation}
\V_{Sch.,1}(r) = \left(1-{2m\over r}\right){\ell(\ell+1)\over r^2} \ , \quad   T(r) = 1-\frac{2m}{r} \ , \quad r_\sigma = 3m \ . 
\end{equation}
Then the relevant quantities necessary to substitute into Eq.~(\ref{perturbativeform}) are:
\begin{eqnarray}
    \V_{Sch.,1}''(r_\sigma) &=& -\frac{2\ell(\ell+1)}{81m^4} \ , \quad \V_{Sch.,1}'''(r_\sigma) = \frac{16\ell(\ell+1)}{243m^5} \ , \nonumber \\
    && \nonumber \\
    T(r_\sigma) &=& \frac{1}{3} \ , \quad T'(r_\sigma) = \frac{2}{9m} \ ,
\end{eqnarray}
and the approximate shift in the QNMs is given by
\begin{eqnarray}
    \delta(\omega^2_n) &\approx& \epsilon\,[\delta\V](r_\sigma) \nonumber \\
    && \qquad +\epsilon\,i\left(n+\frac{1}{2}\right)\left\lbrace\frac{11m}{\sqrt{\ell(\ell+1)}}[\delta\V]'(r_\sigma)-\frac{3m^2}{2\sqrt{\ell(\ell+1)}}[\delta\V]''(r_\sigma)\right\rbrace\Bigg\vert_{r_\sigma=3m} \ . \nonumber \\
    &&
\end{eqnarray}

%========================================================
\section{Conclusions}\label{S:discussion}
%========================================================

The spin-dependent Regge--Wheeler potentials for the regular black hole with asymptotically Minkowski core were extracted and their qualitative features thoroughly analysed. Subsequently, the spin one and spin zero fundamental quasi-normal mode profiles were examined \emph{via} first-order WKB approximation for the respective dominant multipole numbers. For small $a$, both scalar spin zero and spin one electromagnetic fluctuations propagating in a regular black hole with asymptotically Minkowski core spacetime were found to have shorter-lived, higher-energy signals than for their Schwarzschild counterparts. Finally, general analysis of perturbation of the Regge--Wheeler potential itself was performed, and a general result presented explicating the associated shift in the QNM profiles under the perturbation $\V(r)\rightarrow\V(r)+\delta\V(r)$ to first-order in $\epsilon$. This general result was then applied to Schwarzschild spacetime.

Future research could include performing these calculations to higher-order in WKB, using the improved version of WKB with Pad\'{e} approximants, comparing the QNM profiles to those extracted using a different method to WKB (say, \emph{e.g.}, time domain integration), and numerical refinement of the approximations. It would also be prudent to extend the analysis to the spin two axial mode. Discovery of a candidate spacetime which is the asymptotically Minkowski analog to Kerr, on which the wave equation separates, would also be of high interest, giving an astrophysically relevant candidate spacetime which \emph{hopefully} possesses a ringdown signal that LIGO/VIRGO or LISA could delineate from Kerr. Also of interest is to explore the QNMs for when the candidate spacetime is modelling a compact massive object which is something other than a black hole; \emph{i.e.} when $a>2m/\e$.

%=====================================================
\section*{Acknowledgements}
%=====================================================

AS was supported by a Victoria University of Wellington PhD scholarship, and was also indirectly supported by the Marsden Fund, \emph{via} a grant administered by the Royal Society of New Zealand.\\
AS would also like to thank Professor Matt Visser for useful conversations and discussions.

%=====================================================
%===================================================== 

%==================================================================
\end{document}